\documentclass[a4paper,12pt]{article}

\usepackage[cp1251]{inputenc}
\usepackage[russian]{babel}
\usepackage{amsmath, amsthm, amsfonts, amssymb, dsfont, color, accents}
\usepackage{srcltx}

\usepackage{graphicx}

\theoremstyle{plain}

\theoremstyle{remark}

\theoremstyle{definition}

\numberwithin{equation}{section}

\begin{document}

\begin{center}
\bigskip {\LARGE Terrestrial Sagnac delay constraining modified gravity
models }

\bigskip

R.Kh. Karimov$^{1,a}$, R.N. Izmailov$^{1,b}$, A.A.Potapov$^{2,c}$ and K.K.
Nandi$^{1,2,d}$,

\bigskip

$^{1}$Zel'dovich International Center for Astrophysics, Bashkir State
Pedagogical University, 3A, October Revolution Street, Ufa 450000, RB, Russia

$^{2}$Department of Physics \& Astronomy, Bashkir State University, 47A,
Lenin Street, Sterlitamak 453103, RB, Russia

$\bigskip $

$^{a}$E-mail: karimov\_ramis\_92@mail.ru

$^{b}$E-mail: izmailov.ramil@gmail.com

$^{c}$E-mail: potapovaa@mail.ru

$^{d}$E-mail: kamalnandi1952@yahoo.co.in

---------------------------------------------------------------

\bigskip \textbf{Abstract}
\end{center}

Modified gravity theories include $f(\mathbf{R})-$gravity models
that are usually constrained by the cosmological evolutionary scenario.
However, it has been recently shown that they can also be constrained by the
signatures of accretion disk around constant Ricci curvature Kerr-$f(\mathbf{%
R}_{0})$ stellar sized black holes. Our aim here is to use another experimental fact, viz., the
terrestrial Sagnac delay to constrain the parameters of specific $f(\mathbf{R%
})-$gravity prescriptions. We shall assume that a Kerr-$f(\mathbf{R}_{0})$ solution
asymptotically describes Earth's weak gravity near its surface. In this
spacetime, we shall study oppositely directed light beams from
source/observer moving on non-geodesic and geodesic circular trajectories
and calculate the time gap, when the beams re-unite. We obtain the \textit{exact }time gap called Sagnac delay in both
cases and expand it to show how the flat space value is corrected by the
Ricci curvature, the mass and the spin of the gravitating source. Under the
assumption that the magnitude of corrections are of the order of residual
uncertainties in the delay measurement, we derive the allowed intervals for
Ricci curvature. We conclude that the terrestrial Sagnac delay can be used to
constrain the parameters of specific $f(\mathbf{R})$ prescriptions. Despite
using the weak field gravity near Earth's surface, it turns out that the
model parameter ranges still remain the same as those obtained from the
strong field accretion disk phenomenon.


\begin{center}
---------------------------------------------------------------

\textbf{1. Introduction}
\end{center}

Modified gravity theories typically include $f(\mathbf{R})-$gravity models that are mostly constrained by cosmological evolutionary scenario (see, e.g., [1]). On the other hand, the work pioneered by P\'{e}rez, Romero and Perez Bergliaffa [2] seems to be the first in the direction of constraining such models using the strong field effect of accretion into stellar sized spinning black holes characterized by the constant Ricci curvature Kerr-$f(\mathbf{R}_{0})$ gravity solution, which is a chargeless case of the more general Kerr-Newman black hole analyzed in [3] for thermodynamic and
stability properties. The Kerr-$f(\mathbf{R}_{0})$ gravity solution formally \textit{resembles} the Carter solution [4] of general relativity but physically very different from it. We want to clarify that, in the present context, the constant Ricci curvature $\mathbf{R}_{0}$ appearing in the Carter metric has nothing to do with the general relativistic cosmological constant ($\Lambda \approx 10^{-56}$ cm$^{-2}$) but can take on any real value. Indeed, a range of real values for the Ricci curvature $\mathbf{R}_{0} $ can be obtained from various theoretical considerations such as the event horizon or stability of circular orbits [2] or by experimental facts as considered here.

For instance, in P\'{e}rez et al. [2], the authors obtained adimensional curvature ranges $(-\infty ,10^{-6}]$ and $[-1.2\times \ 10^{-3},6.67\times\ 10^{-4}]$ in the cases of Schwarzschild and Kerr black holes in $f(\mathbf{R}_{0})$ gravity, respectively. The lower limit in the latter appear since accretion disk observations of Cygnus-X-1 in the soft state rule out curvature values below $-1.2\times \ 10^{-3}$. These ranges of curvature in turn fix the parameters of specific $f(\mathbf{R})$ prescriptions.

The present work aims to constrain the $f(\mathbf{R})-$gravity prescriptions exploiting a hardcore experimental fact, viz., the residual uncertainty in the terrestrial Sagnac delay observations in 1971 and 1985 experiments. For this purpose, we shall assume that the\ weak gravitational field of spinning Earth near its surface is asymptotically described by the Kerr - $f(\mathbf{R}_{0})$ gravity solution or Carter solution [3]. We shall therefore first calculate the exact Sagnac delay for non-geodesic and geodesic source/observer circular orbits on the equatorial plane of the Earth and then make Parametric Post Newtonian expansions. Our strategy is to identify the observed residual uncertainties in the delay with the leading order corrections to the delay that depend on the local Ricci curvature. This would provide the desired ranges of $\mathbf{R}_{0}$, which in turn would be used to constrain the specific $f(\mathbf{R})$ prescriptions.

The merit of the present work is that the Sagnac delay observations (from Hafele-Keating experiment [5]) in the weak gravity of horizonless Earth corresponding to low Ricci curvature can lead to the same constraints on the $f(\mathbf{R})$ gravity prescriptions as are obtained from strong gravity accretion disk around Kerr-$f(\mathbf{R}_{0})$ gravity black holes.

Hafele and Keating in their historic around-the-Earth\ 1971 experiment carried portable atomic clocks circumnavigating the Earth, once eastward and again westward, and confirmed the special relativistic Sagnac effect attributable to Earth's spin. The clocks will have equal energies but not equal time rates leading to a synchronization discontinuity between them when they reunite. Schlegel [6] has shown that this discontinuity is exactly the same as the light synchronization discontinuity or by another name, the Sagnac delay. Allan, Weiss and Ashby [7] reported in 1985 an equivalent 90 day run of the around-the-world relativistic Sagnac experiment with electromagnetic signals transmitted by GPS satellites, which directly tested the light synchronization discontinuity more accurately. However, even though the delay is caused by spinning Earth, it is not a mass or spin dependent effect to the observed leading order.

A laboratory simulation of the effect is as follows. Consider a circular turntable of radius $R$ having a light source/receiver (meaning the source \textit{and} the receiver at the same point) fixed on the turntable. A beam of light split into two at the source/receiver are made to follow the same closed path along the rim in opposite directions before they are re-united at the source/receiver. If the turntable is not rotating, the beams will arrive at the same time at the source/receiver and an interference fringe will appear. When the turntable rotates with angular velocity $\omega $, the arrival times at the source/receiver will be different for co-rotating and counter-rotating beams: longer in the former case and shorter in the latter. This difference in arrival times is called the flat space Sagnac delay (named after the discoverer), which to leading order in $\omega $ is:
\begin{equation}
\delta \tau _{S}=\frac{4\mathbf{\omega\cdot S}}{c^{2}},
\end{equation}%
where $\mathbf{S}$ is the area of the projection, orthogonal to the rotation axis,\ of the closed path followed by the waves contouring the turntable, $c$ is the speed of light in vacuum and $\mathbf{\omega }$ is the angular velocity of the turntable. It is possible to move ahead from special relativity and consider Kerr-$f(\mathbf{R}_{0})$ corrections to the Sagnac delay (1) due to mass and rotation, when the "turntable" is assumed to be a massive rotating compact object like the Earth. The effect has been previously investigated using different solutions of Einstein's general relativity (see, e.g., [8-13]).

The paper is organized as follows: In Sec.2, we briefly describe the $f(\mathbf{R})-$gravity equations and their solution for a massive spinning compact object. In Sections 3-5, we shall compute the mass and rotation induced corrections to flat space Sagnac delay for circular motion on the equator. \ In Sec.6, we shall derive constraints on $\mathbf{R}_{0}$ from residual error and show in Sec.7 how the ranges of $\mathbf{R}_{0}$ constrain the parameters in two illustrative examples of $f(\mathbf{R})-$ gravity. Sec.8 concludes the paper. We shall choose units such that $16\pi G=c=1$ unless specifically restored.

\begin{center}
\textbf{2. }$f(\mathbf{R})$\textbf{\ gravity equations and }Kerr-$f(\mathbf{R%
}_{0})$\textbf{\ solution}
\end{center}

The $f(\mathbf{R})-$gravity action generalizes Einstein-Hilbert action $S[g]=\int \mathbf{R}\sqrt{-g}d^{4}x$ to
\begin{equation}
S[g]=\int [\mathbf{R}+f(\mathbf{R})+L_{\mathrm{matt}}]\sqrt{-g}d^{4}x,
\end{equation}%
where $g$ is the determinant of the metric tensor and $f(\mathbf{R})$ is an arbitrary function of $\mathbf{R}$. In the metric formalism, the field equations can be obtained by varying the metric, which yields (see, e.g., [2,3]):%
\begin{equation}
\mathbf{R}_{\mu \nu }[1+f^{\prime }(\mathbf{R})]-\frac{1}{2}\left[ \mathbf{R}+f(\mathbf{R})\right] g_{\mu \nu }+\left[ \nabla _{\mu }\nabla _{\nu}-g_{\mu \nu }\Box \right] f^{\prime }(\mathbf{R})+\mathbf{T}_{\mu \nu}=0,
\end{equation}%
where $\mathbf{R}_{\mu \nu }$ is the Ricci tensor, $\Box \equiv \nabla_{\beta }\nabla ^{\beta }$ and $f^{\prime }(\mathbf{R})\equiv df(\mathbf{R})/d\mathbf{R}$ and the stress tensor is defined by
\begin{equation}
\mathbf{T}_{\mu \nu }=\frac{-2}{\sqrt{-g}}\frac{\delta \left( \sqrt{-g}L_{\mathrm{matt}}\right) }{\delta g^{\mu \nu }},
\end{equation}%
where $L_{\mathrm{matt}}$ is the matter Lagrangian. Taking trace of Eq.(3), we get%
\begin{equation}
\mathbf{R}[1+f^{\prime }(\mathbf{R})]-2\left[ \mathbf{R}+f(\mathbf{R})\right]-3\Box f^{\prime }(\mathbf{R})+\mathbf{T}=0.
\end{equation}%
Eqs.(3) are a system of fourth-order nonlinear equations in $g_{\mu \nu }$. In the case of constant Ricci scalar $\mathbf{R}\equiv \mathbf{R}_{0}$, and
in vacuum $\mathbf{T}_{\mu \nu }=0$, Eq.(3) reduces to%
\begin{equation}
\mathbf{R}_{\mu \nu }=\Lambda g_{\mu \nu },
\end{equation}%
\begin{equation}
\Lambda =\frac{f(\mathbf{R}_{0})}{f^{\prime }(\mathbf{R}_{0})-1},
\end{equation}%
and by Eq.(5), $\mathbf{R}_{0}$ satisfies
\begin{equation}
\mathbf{R}_{0}=\frac{2f(\mathbf{R}_{0})}{f^{\prime }(\mathbf{R}_{0})-1}.
\end{equation}%
By appearance, Eq.(6) \textit{looks like} Einstein's general relativity equations with a fixed cosmological constant $\Lambda $ but this similitude is merely notational. As emphasized previously, in the present context the curvature $\mathbf{R}_{0}$, and thus $\Lambda $, can take on arbitrary real values depending on the imposed local physical criteria.

In view of Eqs.(6-8), giving $\Lambda \equiv \mathbf{R}_{0}/2$, the spinning Carter solution [4] can be interpreted as $f(\mathbf{R})-$gravity solution with constant Ricci curvature $\mathbf{R}_{0}$, which reads
\begin{equation}
d\tau ^{2}=\frac{\Delta _{r}}{\rho ^{2}\Xi ^{2}}\left[ dt-a\sin^{2}{%
\theta }d\phi \right] ^{2}-\frac{\Delta _{\theta }\sin^{2}{\theta }}{%
\rho ^{2}\Xi ^{2}}\left[ (r^{2}+a^{2})d\phi -adt\right] ^{2}-\frac{\rho ^{2}%
}{\Delta _{r}}dr^{2}-\frac{\rho ^{2}}{\Delta _{\theta }}d\theta ^{2},
\end{equation}%
where, for convenience, we have written $\gamma \equiv \frac{\mathbf{R}_{0}}{12}>0$, so that
\begin{equation}
\Delta _{r}=(r^{2}+a^{2})\left( 1-\gamma r^{2}\right) -2Mr,
\end{equation}%
\begin{equation}
\rho ^{2}=r^{2}+a^{2}\cos^{2}{\theta },
\end{equation}%
\begin{equation}
\Delta _{\theta }=1+\gamma a^{2}\cos^{2}{\theta },
\end{equation}%
\begin{equation}
\Xi =1+\gamma a^{2},
\end{equation}%
where $M$ is the (asymptotic) mass of the source, $a$ is the ratio between the angular momentum $J$ and the mass $M,$
\begin{equation}
a=\frac{J}{M}.
\end{equation}%
When $\gamma =0$, one recovers the usual Kerr solution of general relativity in Boyer-Lindquist coordinates. We shall compute the Sagnac delay for two
types of equatorial orbits in the ensuing sections and consider only the weak field effects from corresponding expansions.

\begin{center}
\textbf{3. Non-geodesic equatorial orbit}
\end{center}

We shall follow the method developed by Tartaglia [11]. Consider that the source/receiver, sending two oppositely directed light beams, is orbiting around a rotating black hole described by metric (9), along a circumference on the equatorial plane $\theta =\pi /2$. Suitably placed mirrors send back to their origin both beams after a circular trip about the rotating central mass. Assume further that source/receiver is orbiting the central mass at a radius $r=R=$ const. Then the metric (9) reduces to
\begin{eqnarray}
d\tau ^{2} &=& \frac{R^{2}-2MR+a^{2}-\gamma R^{2}(R^{2}+a^{2})}{R^{2}(1+a^{2}\gamma )^{2}}(dt-ad\phi )^{2} \nonumber\\
&& -\frac{1}{R^{2}(1+a^{2}\gamma )^{2}}[(R^{2}+a^{2})d\phi -adt]^{2}.
\end{eqnarray}%
Assuming uniform rotation, the rotation angle $\phi _{0}$ of the source/receiver is
\begin{equation}
\phi _{0}=\omega _{0}t.
\end{equation}%
Since this $\omega _{0}$ is not required to satisfy Kepler's third law, the motion is non-geodesic (see Sec.5). Using $d\phi =d\phi _{0}=\omega _{0}dt$ in Eq.(15), we obtain
\begin{equation}
d\tau ^{2}=\frac{R^{2}[1-(R^{2}+a^{2})\{\omega _{0}^{2}+(a\omega_{0}-1)^{2}\gamma \}]-2MR(a\omega _{0}-1)^{2}}{R^{2}(1+a^{2}\gamma )^{2}}dt^{2}.
\end{equation}%
For light moving along the same circular path it must obey $d\tau =0$. Assuming $\Omega $ to be the angular velocity of light motion along the path, we have
\begin{equation}
R^{2}[1-(R^{2}+a^{2})\{\Omega ^{2}+(a\Omega -1)^{2}\gamma \}]-2MR(a\Omega
-1)^{2}=0, \quad a^{2}\gamma \neq -1.
\end{equation}%
Solving Eq.(18), one finds two roots that represent the angular velocity $\Omega _{\pm }$ of light for the co- and counter rotating light motion given by
\begin{eqnarray}
\Omega _{\pm } &=& \frac{2aM/R+a(R^{2}+a^{2})\gamma }{R^{2}+2(M/R)a^{2}+a^{2}%
\left\{ 1+(R^{2}+a^{2})\gamma \right\} } \nonumber\\
&& \pm \frac{\sqrt{R^{2}-2MR+a^{2}-R^{2}(R^{2}+a^{2})\gamma }}{%
R^{2}+2(M/R)a^{2}+a^{2}\left\{ 1+(R^{2}+a^{2})\gamma \right\} }.
\end{eqnarray}%
The rotation angles $\phi _{\pm }$ for light are then
\begin{equation}
\phi _{\pm }=\Omega _{\pm }t.
\end{equation}%
Eliminating $t$ between Eqs.(16) and (20), we obtain
\begin{equation}
\phi _{\pm }=\frac{\Omega _{\pm }}{\omega _{0}}\phi _{0}.
\end{equation}%
The first intersection of the world lines of the two light rays with the one of the orbiting source/receiver after the emission at time $t=0$ is, when the angles are
\begin{equation}
\phi _{+}=\phi _{0}+2\pi ,
\end{equation}%
\begin{equation}
\phi _{-}=\phi _{0}-2\pi ,
\end{equation}%
which give

\begin{equation}
\frac{\Omega _{\pm }}{\omega _{0}}\phi _{0}=\phi _{0}\pm 2\pi .
\end{equation}%
Solving for $\phi _{0}$, we have
\begin{equation}
\phi _{0\pm }=\mp \frac{2\pi \omega _{0}}{\Omega _{\pm }-\omega _{0}}.
\end{equation}%
Putting the expressions from (19), we obtain
\begin{eqnarray}
\phi _{0\pm }&=&\mp 2\pi \omega _{0}/\left[ \frac{2aM/R+a(R^{2}+a^{2})\gamma }{%
R^{2}+2(M/R)a^{2}+a^{2}\{1+(R^{2}+a^{2})\gamma \}}\right. \nonumber\\
&&\left. \pm \frac{\sqrt{R^{2}-2MR+a^{2}-R^{2}(R^{2}+a^{2})\gamma }}{%
R^{2}+2(M/R)a^{2}+a^{2}\left\{ 1+(R^{2}+a^{2})\gamma \right\} }-\omega _{0}%
\right] .
\end{eqnarray}
The proper time at the rotating source/receiver, deduced from Eq.(17) using Eq.(16), is
\begin{equation}
d\tau =\frac{\sqrt{R^{2}[1-(R^{2}+a^{2})\{\omega _{0}^{2}+(a\omega
_{0}-1)^{2}\gamma \}]-2MR(a\omega _{0}-1)^{2}}}{R(1+a^{2}\gamma )}\frac{%
d\phi _{0}}{\omega _{0}}.
\end{equation}%
Finally, integrating between $\phi _{0-}$ and $\phi _{0+}$ , we obtain the exact Sagnac delay
\begin{equation}
\delta \tau =\frac{\sqrt{R^{2}[1-(R^{2}+a^{2})\{\omega _{0}^{2}+(a\omega
_{0}-1)^{2}\gamma \}]-2MR(a\omega _{0}-1)^{2}}}{R(1+a^{2}\gamma )}\frac{\phi
_{0+}-\phi _{0-}}{\omega _{0}}.
\end{equation}%
Using the integration limits from Eq.(26), we explicitly write it out as
\begin{eqnarray}
\delta \tau =\frac{4\pi }{R}\left[ \left\{
R^{3}+2Ma^{2}+a^{2}R+a^{2}R(R^{2}+a^{2})\gamma \right\} \omega
_{0}-2Ma\right. \nonumber\\
\left. -aR(R^{2}+a^{2})\gamma \right] /\left[ (1+a^{2}\gamma
)\{1-2M/R+4a(M/R)\omega _{0}\right. \nonumber\\
 \left. -(R^{2}+2Ma^{2}/R+a^{2})\omega _{0}^{2}-(a\omega
_{0}-1)^{2}(R^{2}+a^{2})\gamma \}^{1/2}\right] .
\end{eqnarray}
Eq.(29) is the exact Sagnac delay and is often interpreted as the gravitational analogue of the Bohm-Aharonov effect [14] although the light beams are not truly moving in the gravitation free space. The best situation for the gravitational Bohm-Aharonov effect is when the light beams are induced to move along a flat space torus (see for details, Semon [15]). Nevertheless, as shown by Ruggiero [16], expression (29) completely agrees with the one of the gravito-electromagnetic Bohm-Aharonov interpretation [17]. For the viewpoint of Bohm-Aharonov quantum interference in general relativity, see [15,18,19].

On the other hand, we can imagine a static source/receiver keeping a fixed position in a coordinate system defined by distant fixed stars ($\omega_{0}=0$). For him, a Sagnac delay will also occur under the condition that $a\neq 0$, given by
\begin{equation}
\delta \tau _{0}=-\frac{8\pi a\{M+\gamma R(a^{2}+R^{2})/2\}}{R(1+a^{2}\gamma
)\sqrt{1-2M/R-(a^{2}+R^{2})\gamma }}.
\end{equation}%
A Post-Newtonian first order approximation for a static observer sending a pair of light beams in opposite directions along a closed triangular circuit, instead of a circle, was worked out by Cohen and Mashhoon [20] and they found the same result as above in that approximation. So what is important is not the shape but the closedness of the orbit.

\begin{center}
\textbf{4. Post-Newtonian expansion}
\end{center}

We obtain that $\delta \tau $ in Eq.(29) is the Sagnac delay for non-geodesic circular equatorial motion. In most cases many terms in this equation are very small allowing Post-Newtonian series approximations, which we do below. Let us first assume that $\beta =\omega _{0}R\ll 1$, and develop Eq.(29) in powers of $\beta $ retaining terms only up to the second order. The result is
\begin{eqnarray}
\delta \tau &\simeq& -\frac{8\pi a\left\{ M+\gamma R(a^{2}+R^{2})/2\right\} }{R(1+a^{2}\gamma )\sqrt{1-2M/R-(a^{2}+R^{2})\gamma }} \nonumber\\
&&+\frac{4\pi \left\{ R^{2}-2MR+a^{2}-R^{2}(a^{2}+R^{2})\gamma \right\} }{R(1+a^{2}\gamma )\left\{ 1-2M/R-(a^{2}+R^{2})\gamma \right\} ^{3/2}}\beta \nonumber\\
&&-\frac{12\pi a\left[ \{M+\gamma R(a^{2}+R^{2})/2\}\{1-2MR+a^{2}/R^{2}-(a^{2}+R^{2})\gamma \}\right] }{R(1+a^{2}\gamma )\left\{ 1-2M/R-(a^{2}+R^{2})\gamma \right\} ^{5/2}}\beta
^{2},
\end{eqnarray}
which displays that the first term is just $\delta \tau _{0}$ of Eq.(30), as expected. Now we perform a successive post-Newtonian approximation in $\varepsilon =M/R\ll 1$ and in $a/R$ $\ll 1$, and using the expression $\delta \tau _{S}=4\omega _{0}S=4\pi \omega _{0}R^{2}=4\pi \beta R$, we obtain the final result
\begin{eqnarray}
\delta \tau &\simeq& \delta \tau _{S}\left\{ 1+\frac{\gamma R^{2}}{2}-\gamma a^{2}\left( 1+\frac{\gamma R^{2}}{2}\right) \right\} \nonumber\\
&&+4\pi RM\omega _{0}\left\{ 1+\frac{3\gamma R^{2}}{2}-\gamma a^{2}\left( 1+\frac{3\gamma R^{2}}{2}\right) \right\} \nonumber\\
&&-\frac{8\pi aM}{R}\left\{ 1+\gamma R^{2}-\gamma a^{2}\left( 1+\gamma R^{2}\right) \right\} .
\end{eqnarray}
This expression reduces to the corresponding one of Tartaglia [11] when $\gamma =0$. Eq.(32) can be re-organized as%
\begin{equation}
\delta \tau \simeq \left( \delta \tau _{S}+4\pi RM\omega _{0}-\frac{8\pi aM}{%
R}\right) + \mathrm{terms}\;\mathrm{dependent}\;\mathrm{on}\;\gamma .
\end{equation}%
The corrections to $\delta \tau _{S}$ due to $\gamma $, $M$ and $a$ are evident. However, the flat space Sagnac effect $\delta \tau _{S}$ is \textit{%
not} completely recovered even when the correction terms containing $M$ and $a$ are negligible, due to the appearance of an extra term $\frac{\gamma R^{2}}{2}$, which comes in as a contribution of constant curvature scalar $\mathbf{R}_{0},$ at an orbit radius of Earth (say) $r=R_{\oplus }$. The
terms proportional to $\left( \frac{\gamma R^{2}}{2}\right) $ or $\left(\mathbf{R}_{0}R^{2}/24\right) $ can be interpreted as an $f(\mathbf{R}_{0})$
contribution, provided it does not vanish. We shall estimate the bounds on\textbf{\ }$\mathbf{R}_{0}$ soon.

\begin{center}
\textbf{5. Geodesic equatorial orbit}
\end{center}

The equatorial orbit in Sec.3 was not geodesic or in free fall since the source/receiver was at the Earth's radius $R_{\oplus }$ sharing its constant rotational velocity $\omega _{0}=\Omega _{\oplus }=7.30\times 10^{-5}$ rad/s, but the motion was not required to satisfy Kepler's third law. Here we are considering a circular geodesic orbit of the source/receiver at some arbitrary radius on the equator ($\theta =\pi /2$) and sending light signals circumnavigating the Earth. The rotational velocity $\omega _{\pm }$ of the satellite is now determined by the circular geodesic itself as follows.

Defining the velocity four-vector $\dot{x}^{\nu }=\frac{dx^{\nu }}{d\tau}$, the Lagrangian can be written as%
\begin{equation}
L=\frac{1}{2}g_{\mu \nu }\dot{x}^{\mu }\dot{x}^{\nu }
\end{equation}%
and the Euler-Lagrange $r-$equation is
\begin{equation}
\frac{d}{d\tau }\left( \frac{\partial L}{\partial \dot{r}}\right) =%
\frac{\partial L}{\partial r}.
\end{equation}%
Since in metric (9), $g_{r\mu }=0$ for $r\neq \mu $, we have%
\begin{equation}
\frac{d}{d\tau }\left( g_{rr}\dot{r}\right) =\frac{1}{2}g_{\mu \nu ,r}%
\dot{x}^{\mu }\dot{x}^{\nu }.
\end{equation}%
Circular orbits are defined by the conditions%
\begin{equation}
\dot{r}=\ddot{r}=0,
\end{equation}%
so that the Eq.(35) yields%
\begin{equation}
g_{tt,r}\dot{t}^{2}+2g_{t\phi ,r}\dot{t}\dot{\phi}+ g_{\phi \phi ,r}\dot{\phi}^{2}=0.
\end{equation}%
Defining $\omega =\dot{\phi}/\dot{t}$, this equation yields the quadratic equation \
\begin{equation}
g_{\phi \phi ,r}\omega ^{2}+2g_{t\phi ,r}\omega +g_{tt,r}=0.
\end{equation}%

From the metric (9), putting $dr=0$ at $r=R=$ const. and $d\theta =0$ at $\theta =\pi /2$, we find%
$$d\tau ^{2}=g_{tt}dt^{2}+2g_{t\phi }dtd\phi +g_{\phi \phi }d\phi ^{2},$$
where%
$$g_{tt} =1-\frac{2M}{R}-\gamma \left( a^{2}+R^{2}\right) ,\quad g_{t\phi}=\frac{2aM}{R}+\gamma a\left( a^{2}+R^{2}\right)$$
\begin{equation}
g_{\phi \phi } =-\frac{2a^{2}M}{R}-\left( a^{2}+R^{2}\right) \left(1+a^{2}\gamma \right) .
\end{equation}%
The source/receiver rotational velocities $\omega _{\pm }$ then follow from the two roots of Eq.(38), using Eqs.(39),
\begin{eqnarray}
\omega _{\pm } &=&\frac{\left( \frac{aM}{R^{2}}-aR\gamma \right) \pm \sqrt{\frac{M}{R}-\gamma R^{2}}}{\frac{a^{2}M}{R^{2}}-R-a^{2}R\gamma }, \nonumber\\
\delta \tau _{S\pm } &=&4\pi R^{2}\omega _{\pm }.
\end{eqnarray}%
The above $\delta \tau _{S\pm }$ is the exact delay for geodesic motion. One could treat this result as representing the effect of cosmological constant $\Lambda $ on the Sagnac delay. When $a=0$, $\gamma =0$, we have $\omega_{\pm }=\mp \sqrt{\frac{M}{R^{3}}}$, which is just Kepler's third law. We now expand Eq.(40) up to first order in $\left( a/R\right) $ and obtain
\begin{equation}
\omega _{\pm }=\left( \gamma R-\frac{M}{R^{2}}\right) \left( \frac{a}{R}%
\right) \pm \frac{1}{R}\sqrt{\frac{M}{R}-\gamma R^{2}}.
\end{equation}%
Noting that $\omega _{\pm }$ $=$ const. (since $r=R=$ const. for circular orbits), we can insert it into the first order delay\ to obtain $\delta \tau_{S\pm }=4\pi R^{2}\omega _{\pm }$, so that
\begin{equation}
\delta \tau _{S\pm }^{\mathrm{geod}}=\pm 4\pi \left[ \sqrt{MR-\gamma R^{4}}\mp
\left( \frac{a}{R}\right) \left( M-\gamma R^{3}\right) \right].
\end{equation}%
The Kerr terms follow at $\gamma =0$, when we recover the formula derived by Lichtenegger and Iorio [21]:
\begin{equation}
\delta \tau _{\pm }=\pm 4\pi \sqrt{MR}\mp 4\pi a\left( \frac{M}{R}\right).
\end{equation}

\begin{center}
\textbf{6. Constraints on }$\mathbf{R}_{0}$ \textbf{from the terrestrial
Sagnac data }
\end{center}

We shall consider that the source/receiver is orbiting along a circular path close to the spinning Earth, assuming that Earth's gravity near its surface (weak field) is described by the Carter metric of Kerr-$f(\mathbf{R}_{0})$ gravity. The gravitational field of the Earth has already been described in the weak field by the Kerr metric leading, for instance, to the Lense-Thirring (LT) precession already well tested by LAGEOS and Gravity Probe - Gravity Probe - B missions [22-24]. Hackmann and L\"{a}mmerzahl [25-27] have recently given an expression of LT precession that is valid up to first order in the Kerr parameter $a$ for a more general axisymmetric six-parameter Pleba\'{n}ski-Demia\'{n}ski spacetime, of which the presently considered $f(\mathbf{R}_{0})$ - Kerr solution is just a special case [25]. See also [26,27] for more complete details.

To obtain bounds on $\mathbf{R}_{0}$ from the terrestrial Sagnac delay, we consider the relevant Earth data:
\begin{eqnarray}
R_{\oplus }&=&6,378,137 \;\mathrm{m}, \nonumber\\
\Omega _{\oplus }&=&7.30\times 10^{-5} \;\mathrm{rad/s}, \nonumber\\
r_{g}&=&GM_{\oplus }/c^{2}=4.35\times 10^{-3} \;\mathrm{m}, \nonumber\\
a &=&a_{\oplus }=9.81\times 10^{6} \;\mathrm{m}^{2}/\mathrm{s}, \nonumber\\
c &=&3\times 10^{8} \;\mathrm{m/s}.
\end{eqnarray}%
The basic total Sagnac delay $\delta \tau _{S}=4\pi \omega _{0}R^{2}/c^{2}$, with $\omega _{0}=\Omega _{\oplus }$, $R=R_{\oplus }$, due to the east and westward equatorial motion of the source/receiver, works out to
\begin{equation}
\delta \tau _{S}=2\times \frac{2\Omega _{\oplus }}{c^{2}}\times \pi
R_{\oplus }^{2}=4.148\times 10^{-7}\mathrm{s}=2\times 207.4 \;\mathrm{nsec}.
\end{equation}%
As well known, this famous value $\frac{1}{2}\delta \tau _{S}$ ($=207.4$ nsec) is the one way delay (either east or westward circumnavigation) compared to a stationary clock on Earth that has been measured (\textit{excluding} velocity and altitude factors) by Hafele and Keating in their famous airborne atomic clock experiment [5]. Schlegel [6] explained that the Hafele-Keating value of clock synchronization discontinuity between the flying equatorial clocks and the stationary clock on Earth is exactly the Sagnac delay $\frac{2\Omega _{\oplus }}{c^{2}}\pi R_{\oplus }^{2}$. We shall now consider various corrections contributing to the\ one way basic delay $\frac{1}{2}\delta \tau $.

\textit{Correction to delay for non-geodesic equatorial motion}

The Hafele-Keating around-the-world experiment involved portable atomic clocks that had undergone non-geodesic equatorial motion because those were propelled by the aircraft engine. Therefore, Eq.(32)\ for total observed delay $\delta \tau $ is applicable and the corrections contributed by $M$, $\omega _{0}$, $\gamma $ and $a$ to the one way basic Sagnac delay $\frac{1}{2}\delta \tau _{S}$ can be obtained by computing $\frac{1}{2}\left( \delta \tau -\delta \tau _{S}\right) $. Putting in the relevant Earth values in the expression (32) for $\delta \tau $, using $\delta \tau _{S}=4.148\times 10^{-7}$s, and restoring $\gamma =\frac{\mathbf{R}_{0}}{12}$, we obtain
\begin{eqnarray}
\Delta\tau_{\mathrm{non-geo}}^{\mathrm{corr}} &=& \frac{1}{2}\left( \delta \tau
-\delta \tau _{S}\right) \\
&=&6.98\times 10^{14}\mathbf{R}_{0}-6.22\times 10^{-14}\mathbf{R}_{0}^{2}.
\end{eqnarray}%
The dependence of correction on the unknown Ricci curvature $\mathbf{R}_{0}$ is evident. In order to have an idea of its possible numerical range, let us consider the underlying metric signature of (9) by putting $a=0$. Then, one has $\Delta _{r}=r^{2}\left( 1-\gamma r^{2}\right) -2Mr$. If $\gamma r^{2}\geq 1$, then $\Delta _{r}<0$, which would imply an invalid metric signature ($t,r,\theta ,\phi $) $\rightarrow $ ($-,+,-,-$). Therefore, one must have
\begin{equation}
0\leq \gamma r^{2}<1,
\end{equation}%
yielding a valid signature ($t,r,\theta ,\phi $) $\rightarrow $ ($+,-,-,-$). This signature must be preserved throughout the spacetime and the inequality (48) should be canonical. Then the constraint (48) reads%
\begin{equation}
0\leq \mathbf{R}_{0}<\frac{12}{R_{\oplus }^{2}}.
\end{equation}%
Using the above inequality, with $R_{\oplus }$ from (44), we get the range $0\leq \mathbf{R}_{0}<2.95\times 10^{-13}$ m$^{-2}$, which must not be violated.

We now assume, as an input, that the correction $\Delta \tau _{\mathrm{non-geo}}^{\mathrm{corr}}$ in (47) is sunk in the observed maximum residual error $\sim 10$ ns [5], i.e., we assume that (using the conversion $1$ s $=10^{9}$ ns):
\begin{equation}
10^{9}\times \Delta \tau _{\mathrm{non-geo}}^{\mathrm{corr}}=10 \;\mathrm{ns},
\end{equation}%
which yields two roots $\mathbf{R}_{0}^{(1)}=1.43\times 10^{-14}$ m$^{-2}$ and $\mathbf{R}_{0}^{(2)}=1.22\times 10^{28}$ m$^{-2}$. The latter root is discarded since we are considering an experiment in the weak field limit in the vicinity of the Earth's surface, where curvature is expected to be extremely low.$a=a_{\oplus }$, Thus, one ends up with a slightly sharper range given by
\begin{equation}
0\leq \mathbf{R}_{0}<1.43\times 10^{-14} \;\mathrm{m}^{-2}.
\end{equation}%
We shall see below that this range can be sharpened further using the geodesic motion.

\textit{Correction to delay for geodesic equatorial motion}

It should be noted that Allan, Weiss and Ashby [7] reported in 1985 an equivalent 90 day run of the around-the-world relativistic Sagnac experiment with electromagnetic signals transmitted by GPS satellites, which directly test the synchronization discontinuity leading to \ reduced residual error of $\sim 5$ ns. Here we are enumerating a delay, where the source/receiver and electromagnetic signals are undergoing geodesic (free fall satellites) motion, unlike in the Hafele-Keating airborne experiment. So, some difference in the range of $\mathbf{R}_{0}$ would be expected here, even if the same circular orbital radius $R_{\oplus }$ is chosen. Recall the one way delay:
$$\left( \frac{1}{2}\right) \delta \tau _{S\pm }^{\mathrm{geo}}=\pm 2\pi \left[ \sqrt{MR-\gamma R^{4}}\right] +\left( \frac{a}{R}\right) \left( M-\gamma R^{3}\right). $$
In order that the delay be not imaginary (reality constraint), the first term in (42) provides a curvature range
\begin{equation}
0\leq \mathbf{R}_{0}<\frac{12GM_{\oplus }}{c^{2}R_{\mathrm{geo}}^{3}},
\end{equation}%
which is specific to geodesic motion.\ Choose an\ approximate orbit radius around the Earth, $R_{\mathrm{geo}}=7\times 10^{6}$ m [11], then the signature constraint (48) immediately gives $0\leq \mathbf{R}_{0}<1.54\times 10^{-22}$ m$^{-2}$. We can try to find the value of\ $\mathbf{R}_{0}$ from the correction term: Restoring $\gamma $, $G$ and $c$, and $R=R_{\mathrm{geo}}$ in Eq.(42), we obtain the one way delay

\begin{eqnarray}
\left( \frac{1}{2}\right) \delta \tau _{S}^{\mathrm{geod}} &\simeq &\left\vert
\frac{2\pi }{c}\sqrt{\frac{GM_{\oplus }R_{\mathrm{geo}}}{c^{2}}}\right\vert %
\left[ 1+\frac{c^{2}\mathbf{R}_{0}R_{\mathrm{geod}}^{3}}{24GM_{\oplus }}\right] \nonumber\\
&=&3670\times \left[ 1+3.25\times 10^{21}\mathbf{R}_{0}\right] \;\mathrm{ns}.
\end{eqnarray}%
A free fall satellite transmitting light in geodesic motion in both directions at the radius $R_{\mathrm{geo}}$ is predicted to measure the basic
two way delay (just double, $2\times 3670$ nsec $=7.34\times 10^{-6}$ sec). Precisely, this is the value also obtained by Tartaglia [11]. As in (46), we
constrain the correction term $\Delta \tau _{\mathrm{geod}}^{\mathrm{corr}}$, that is, the second term in Eq.(53), by $0\leq \Delta \tau_{\mathrm{geod}}^{\mathrm{corr}}\leq $ the observed error residual, to find
\begin{equation}
\Delta \tau _{\mathrm{geod}}^{\mathrm{corr}}=1.2\times 10^{25}\mathbf{R}_{0}%
\mathrm{ns}\Rightarrow \mathbf{R}_{0}=\frac{\Delta \tau _{\mathrm{geod}}^{\mathrm{corr}}}{1.2}\times 10^{-25}\;\mathrm{m}^{-2}\left(\mathrm{ns}\right) ^{-1}.
\end{equation}%
Experiments involving geodesic motion of clocks in circular orbit have not been done. Nevertheless, taking into account the refined error residual $%
\sim 5$ ns\ [7] to the basic delay $3670$ nsec, the corresponding range of $\mathbf{R}_{0}$ becomes\footnote{Even if the error residual is a bit higher, it does not significantly alter the limit on $\mathbf{R}_{0}$.}%
\begin{eqnarray}
\Delta \tau _{\mathrm{geod}}^{\mathrm{corr}}=1.2\times 10^{25}\mathbf{R}_{0}\;\mathrm{ns}\sim 5\;\mathrm{ns} \nonumber\\
\Rightarrow 0\leq \mathbf{R}_{0}\leq 4.16\times 10^{-25}\;\mathrm{m}^{-2},
\end{eqnarray}
which we argue to be the range associated with the free fall geodesic motion that is seen to be finer than the previous range (51) by over ten orders of magnitude. Incidentally, Ruggiero [24] has shown how the presence of $\mathbf{R}_{0}$ affects the Kepler rotational velocity, precession of the pericenter and the angular velocity of the gravitomagnetic precession. The accuracy achieved in the Gravity-Probe-B experiment, completed several years ago, is $7.2$ milliarcseconds/year\footnote{We thank an anonymous reviewer for pointing it out to us.} and it is possible to deduce an estimate\footnote{Ruggiero's notation $k$ is the same as $\mathbf{R}_{0}/4$ or $\Lambda /2$ in our notation.} $\left\vert k\right\vert $ $\leq 10^{-26}$ m$^{-2}$, which is quite near the upper limit derived in (56). We shall consider only the interval (55) for our computation below as\ it is a bit finer.

\begin{center}
\textbf{7. Illustrative examples of }$f(\mathbf{R})-$\textbf{gravity}
\end{center}

The examples of specific $f(\mathbf{R})-$gravity models and their analyses are patterned after those in P\'{e}rez et al. [2]. Since the Sagnac delay
allows only a positive range of $\mathbf{R}_{0}$ as in (56), we do not discuss negative values of $\mathbf{R}_{0}$ in what follows.

(1) Consider the model
\begin{equation}
f(\mathbf{R})=\alpha \mathbf{R}^{\beta },
\end{equation}%
where the constants $\alpha $, $\beta $ and the constant Ricci scalar $%
\mathbf{R}_{0}$ are related by Eq.(8) as
\begin{equation}
\mathbf{R}_{0}=\left[ \frac{1}{\alpha \left( \beta -2\right) }\right] ^{\frac{1}{\beta -1}}.
\end{equation}%
Note that $\beta >0$ and small positive values of $\alpha >0$ are necessary conditions that ensure the passage to general relativity for small values of the Ricci scalar $\mathbf{R}$. With $r_{g}=GM_{\oplus}/c^{2}=4.35\times 10^{-3}$ m, and the adimensional curvature\ defined as
\begin{equation}
\mathbb{R}_{0}=\mathbf{R}_{0}r_{g}^{2},
\end{equation}%
the range (55), $0\leq \mathbf{R}_{0}\leq 4.16\times 10^{-25}$ m$^{-2}$, now translates to the adimensional range
\begin{equation}
0\leq \mathbb{R}_{0} < 7.87\times 10^{-30}.
\end{equation}%
Further, with $R_{g}=r_{g}^{2}$, the parameter $\alpha $ is redefined as $\alpha ^{\prime }=\alpha R_{g}^{\beta -1}$ so that one can rewrite Eq.(57)
as
\begin{equation}
\mathbb{R}_{0}=\left[ \frac{1}{\alpha ^{\prime }\left( \beta -2\right) }\right] ^{\frac{1}{\beta -1}}\Rightarrow \alpha ^{\prime }\left( \beta ,\mathbb{R}_{0}\right) =\frac{1}{\mathbb{R}_{0}^{\beta -1}}\left[ \frac{1}{\beta -2}\right].
\end{equation}%
It is clear from the above that $\beta =2$ is ruled out because the Eq.(60)\ is not defined and $\beta >2$ is ruled out as it leads to very large values
of $\alpha ^{\prime }$, hence of $\alpha $, thereby preventing the passage to general relativity.
\begin{figure}
  \includegraphics[scale=0.5]{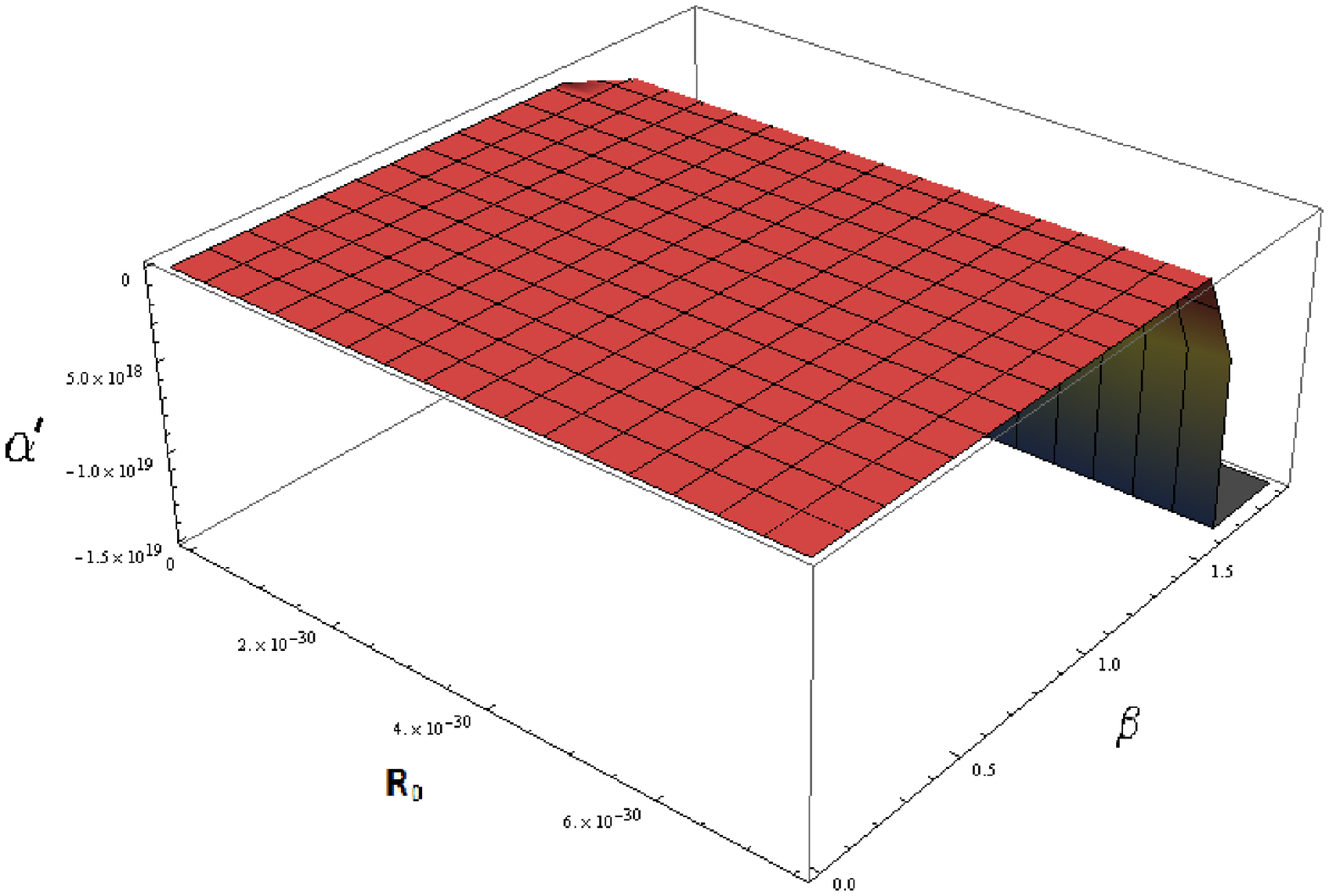}
\caption{Plot of $\protect\alpha^{\prime }\left(\protect\beta, \mathbb{R}_{0}\right)$ as a function of $\protect\beta >0$ and $\mathbb{R}_{0}$.}
\label{fig:1}
\end{figure}
From the Fig.1, within the interval $\beta \in (0,2)$, we then find that $\alpha ^{\prime }\in (-\infty ,0)$ for the range $0\leq \mathbb{R}_{0}<7.87\times 10^{-30}$, which leads to the following constraints on the parameters of the model
\begin{equation}
\alpha ^{\prime }\in (-\infty ,0),\quad \beta \in (0,2),\quad \mathbb{R}_{0}\in (0,7.87\times 10^{-30}].
\end{equation}%
These constraints are further modified when the generic viability conditions of $f(\mathbf{R})-$gravity are imposed. These are (see, e.g., [1])%
\begin{equation}
-1<f^{\prime }(\mathbf{R}_{0})<0,
\end{equation}
\begin{equation}
f^{\prime \prime }(\mathbf{R}_{0})>0.
\end{equation}%
The first condition ensures an effective positive gravitational constant and the second condition is necessary to avoid the Dolgov-Kawasaki [28]
instability of the Ricci scalar. The two conditions respectively yield%
\begin{equation}
-1<\alpha \beta \mathbf{R}_{0}^{\beta -1}<0,
\end{equation}
\begin{equation}
\alpha \beta (\beta -1)\mathbf{R}_{0}^{\beta -2}>0.
\end{equation}%
In view of the second part of (65) giving $\alpha <0$, the inequality (66) yields
\begin{equation}
0<\beta <1.
\end{equation}%
Because of the inequality (66), we can write using the first part of (65) another valid inequality:%
\begin{equation}
\alpha \mathbf{R}_{0}^{\beta -1}>-1\Rightarrow \alpha >-\frac{1}{\mathbf{R}_{0}^{\beta -1}}.
\end{equation}%
The following are the limits on $\alpha $ induced by the two limiting values of $\beta $. If $\beta =0$, $\alpha >-\mathbf{R}_{0}$ and for $\beta =1$, $\alpha >-1$. Together with $\alpha <0$ from (65), the range for $\alpha $ is then $\alpha \in (-\mathbf{R}_{0},0).$ Thus the final range of parameters finally constrained by the viability/instability conditions is%
\begin{equation}
\alpha \in (-\mathbf{R}_{0},0),\beta \in (0,1),\mathbb{R}_{0}\in (0,7.87\times 10^{-30}].
\end{equation}%
The first two intervals are the same as in [2], while the curvature interval is much smaller representing the weak field of the Earth.

(2) The chosen function is

\begin{equation}
f(\mathbf{R})=\epsilon \mathbf{R}\ln \frac{\mathbf{R}}{\alpha },
\end{equation}%
where the parameters $\epsilon $ and $\alpha $ and the constant Ricci scalar $\mathbf{R}_{0}$ are related by Eq.(8) as
\begin{equation}
\alpha =\mathbf{R}_{0}\exp \left( \frac{1}{\epsilon }-1\right) .
\end{equation}%
Adimensionalizing as before, we have
\begin{equation}
\alpha ^{\prime }\left( \epsilon ,\mathbb{R}_{0}\right) =\mathbb{R}_{0}\exp \left( \frac{1}{\epsilon }-1\right) .
\end{equation}

\begin{figure}
  \includegraphics[scale=0.5]{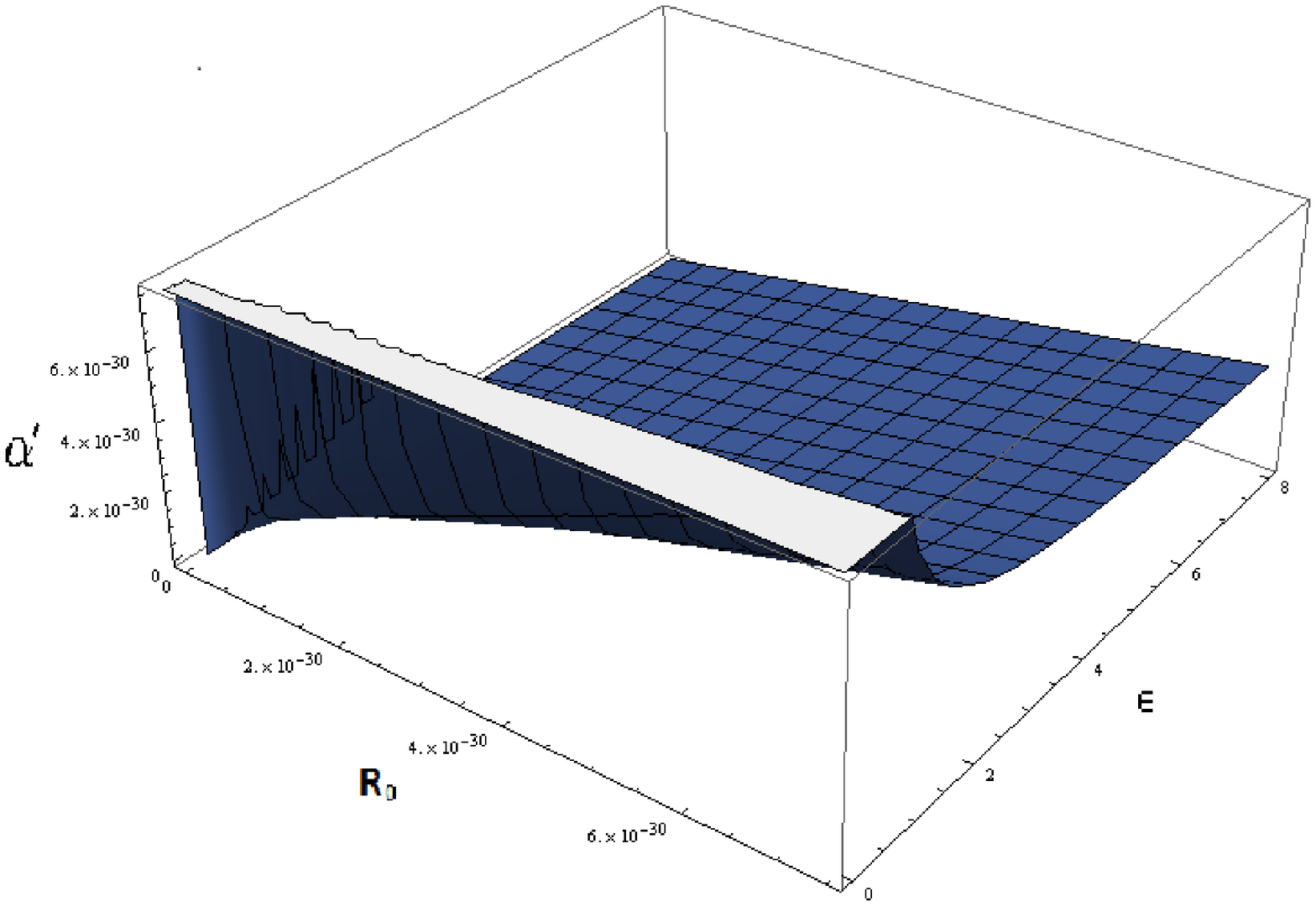}
\caption{Plot of $\protect\alpha^{\prime }\left( \protect\epsilon ,\mathbf{R}_{0}\right) $ as a function of $\protect\epsilon \in (0,8)$ and $\mathbf{R}_{0}$.}
\label{fig:2}
\end{figure}
\begin{figure}
  \includegraphics[scale=0.5]{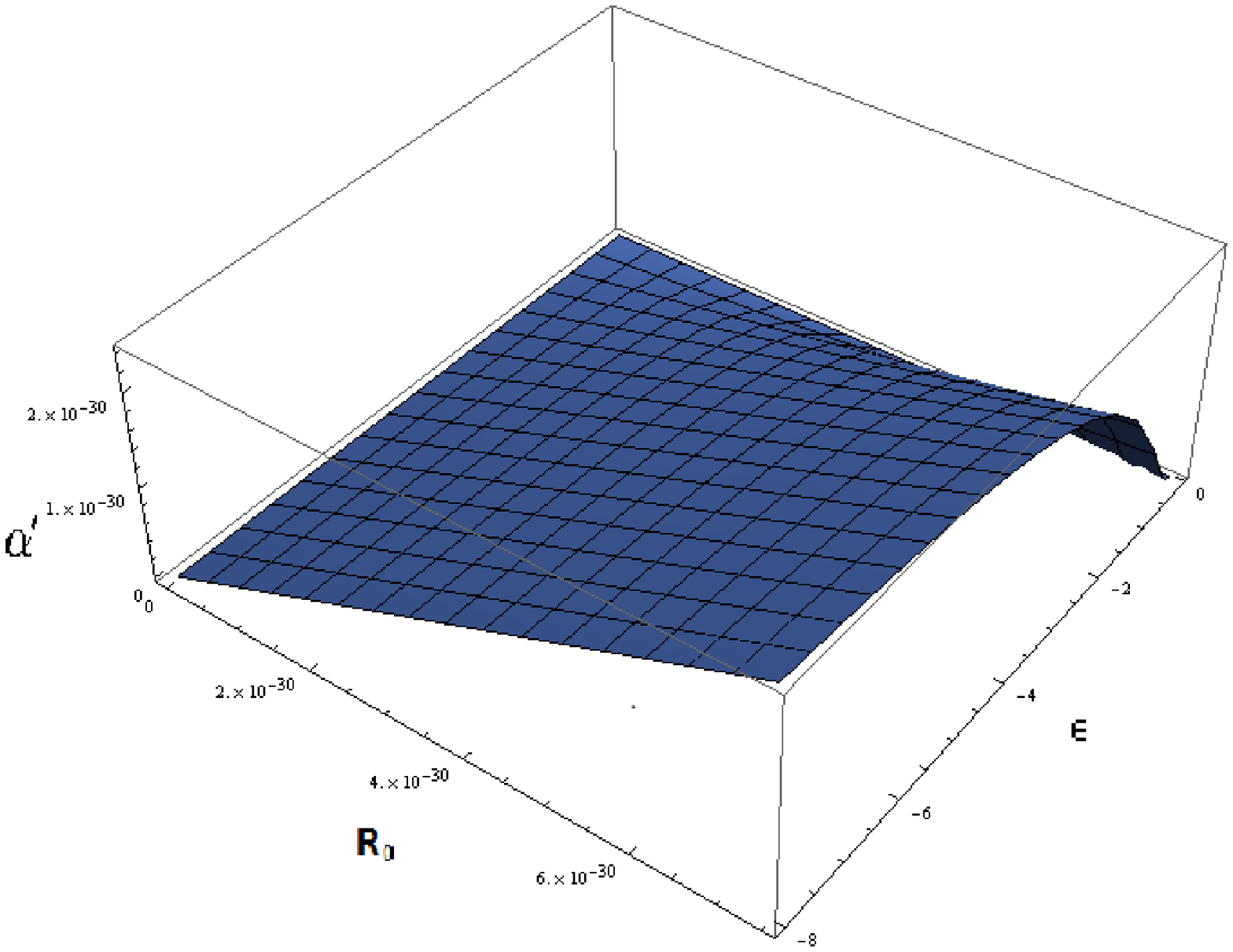}
\caption{Plot of $\protect\alpha^{\prime }\left( \protect\epsilon ,\mathbb{R}_{0}\right) $ as a function of $\protect\epsilon \in (-8,0)$ and $\mathbb{R}_{0}$.}
\label{fig:3}
\end{figure}
From Eq.(70) and as illustrated in Figs.2 and 3, it follows that two cases are possible for $\mathbf{R}_{0}>0$:%
\begin{equation}
\epsilon \in (-\infty ,0)\quad\Rightarrow \alpha ^{\prime }\in \left(0,e^{-1}\mathbb{R}_{0}\right),
\end{equation}
\begin{equation}
\epsilon \in (0,\infty )\quad\Rightarrow \alpha ^{\prime }\in \left(e^{-1}\mathbb{R}_{0},\infty \right).
\end{equation}%
The viability condition (63) and the stability condition (64) then respectively imply
\begin{equation}
-1<\epsilon \left( 1+\ln \frac{\mathbb{R}_{0}}{\alpha ^{\prime }}\right) <0,
\end{equation}
\begin{equation}
\frac{\epsilon }{\mathbb{R}_{0}}>0.
\end{equation}%
For $\mathbf{R}_{0}>0$, as suggested by the Sagnac delay data, the only possibility from the last inequality is that $\epsilon >0$, and from the inequality (75), it follows that $\alpha ^{\prime }\in \left(e^{-1}\mathbb{R}_{0},\infty \right) $. Summarizing, the constraints on the parameters of this model are
\begin{equation}
\epsilon >0, \quad \alpha ^{\prime }\in \left( e^{-1}\mathbb{R}_{0},\infty \right),\quad\mathbb{R}_{0}\in (0,7.87\times 10^{-30}].
\end{equation}%
The above two examples show that the free parameters of the models are constrained to remain in the same interval although the interval for
curvature is considerably much smaller than that corresponding to strong curvature accretion disk phenomenon [2], which is $\mathbb{R}_{0}\in \lbrack -1.2\times 10^{-3},6.67\times 10^{-4}]$.

\begin{center}
\textbf{8. Conclusions}
\end{center}

The terrestrial Sagnac effect being investigated in the $f(\mathbf{R}_{0})$ -- Kerr gravity in this paper is probably the first of its kind, to our knowledge. The spinning source introduces a synchronization discontinuity (interpreted as the Sagnac effect [5-7]) between two oppositely directed light beams when they are re-united. Three main conclusions of the work are as follows:

(1) Eqs.(29) and (41) for the \textit{exact} Sagnac delay in the $f(\mathbf{R}_{0})$ - Kerr gravity are the pivotal results of this paper. Approximations have been made to expose, in Eqs.(32) and (43), the leading order corrections due to $M,a,\omega _{0}$ and $\mathbf{R}_{0}$ to flat space value $\delta \tau _{S}$ measured in the weak field Hafele-Keating experiment.

(2) For our purposes, we do not need the basic terrestrial value $\delta\tau _{S}$ but need only the corrections thereto. Using the more refined
error residual, together with the input assumption that the correction $\Delta \tau _{\mathrm{geod}}^{\mathrm{corr}}$ is sunken in the error residual,
viz., $0\leq \Delta \tau _{\mathrm{geod}}^{\mathrm{corr}}\leq $ error residual $\sim 5$ ns, we end up with the range (56), viz., $0\leq \mathbf{R}_{0}\leq 4.16\times 10^{-25}$ m$^{-2}$. Since it is sharper than that from non-geodesic motion, it is used to fix parameters of specific $f(\mathbf{R})- $gravity models.

(3) P\'{e}rez et al. [2] obtained the range for $\mathbb{R}_{0}$ using the accretion disk phenomenon in the strong field of the X-ray binary Cygnus X-1 black hole, which yielded an adimensional $\mathbb{R}_{0}\in \lbrack -1.2\times 10^{-3},6.67\times 10^{-4}]$. It follows that our obtained range in (56), viz., in $\mathbb{R}_{0}\in (0,7.87\times 10^{-30}]$ from the terrestrial scenario is much smaller, which is expected, since in the vicinity of Earth's surface, gravity is weak. Following their analysis, the parameters of the same $f(\mathbf{R})-$gravity models have been constrained by (69)\ and (77), which are found to be the same as those in P\'{e}rez et al. [2] despite the fact that our circular motions lie in the weak field of spinning \textit{horizonless} Earth.

As a curiousity, since the field equations (6) for constant Ricci scalar $f(\mathbf{R})$ -- gravity anyway resemble Kerr-de Sitter equations, one could
identify $\mathbf{R}_{0}=2\Lambda $ and compare the limits on cosmological constant $\Lambda $ obtained in the literature. Kagramanova, Kunz and L\"{a}%
mmerzahl [29] derived several limits ranging from $\left\vert \Lambda \right\vert \leq $ $6\times 10^{-24}$ m$^{-2}$ (gravitational time delay) to
$\left\vert \Lambda \right\vert \leq $ $10^{-41}$ m$^{-2}$ (perhelion shift). Similarly, Sereno and Jetzer [30] considered observations that cover
distance scales between $\sim 10^{8}$ to $\sim 10^{15}$ km and showed that the best constraint $\left\vert \Lambda \right\vert \leq $ $10^{-42}$ m$%
^{-2} $ comes from the perihelion precessions of Earth and Mars, a conclusion reached also in [29]. All the limits are evidently far larger
than the cosmology motivated value $\Lambda \sim 10^{-52}$ m$^{-2}$ and it seems hopeless to try to determine it from local effects. Nonetheless, our
limit $\left\vert \Lambda \right\vert =\frac{\mathbf{R}_{0}}{2}\leq 2.08\times 10^{-25}$ m$^{-2}$ comes closer to that obtained by Kagramanova
et al. [29] using the gravitational time delay experiment as well as to that argued by Ruggiero [24], viz., $\left\vert \Lambda \right\vert \leq $ $%
10^{-26}$ m$^{-2}$, on the basis of high accuracy of the Gravity Probe-B experiment.

\textbf{Acknowledgment}

Part of the work was supported by the Russian Foundation for Basic Research
(RFBR) under Grant No.16-32-00323.

\textbf{References}

[1] S. Capozziello and V. Faraoni, \textit{Beyond Einstein Gravity: A Survey
of Gravitational Theories for Cosmology and Astrophysics,} Fundamental
Theories of Physics 170 (Springer, New York, 2011).

[2] D. P\'{e}rez, G. E. Romero, and S.E. Perez Bergliaffa, Astron.
Astrophys. \textbf{551}, A4 (2013).

[3] J.A.R. Cembranos, A. de la Cruz-Dombriz, P. Jimeno Romero, Int. J. Geom.
Meth. Mod. Phys. \textbf{11}, 1450001 (2014).

[4] B. Carter, in \textit{Les Astres Occlus} eds. C. DeWitt, \& B. DeWitt
(Gordon \& Breach, New York, 1973).

[5] J.C. Hafele and R.E. Keating, Science \textbf{177}, 166 (1972), \textit{%
ibid. }\textbf{177}, 168 (1972).

[6] R. Schlegel, Nature (London), \textbf{242}, 180 (1973).

[7] D.W. Allan, M.A. Weiss and N. Ashby, Science \textbf{228}, 69 (1985

[8] A. Bhadra, T.B. Nayak and K.K. Nandi, Phys. Lett. A\textbf{\ 295}, 1
(2002).

[9] K.K. Nandi, P.M. Alsing, J.C. Evans and T.B. Nayak, Phys. Rev. D\textbf{%
\ 63}, 084027 (2001).

[10] A. Ashtekar and A. Magnon, J. Math. Phys. \textbf{16}, 343 (1975).

[11] A. Tartaglia, Phys. Rev. D\textbf{\ 58}, 064009 (1998).

[12] J. Sultana, Gen. Rel. Grav. \textbf{46}, 1710 (2014).

[13] B. Mashhoon, Phys. Lett. A \textbf{173}, 347 (1993).

[14] Y. Aharonov and D. Bohm, Phys. Rev. \textbf{115}, 485 (1959).

[15] M. D. Semon, Found. Phys. \textbf{12}, 49 (1982).

[16] M. L. Ruggiero, Nuovo Cim. B \textbf{119}, 893 (2004).

[17] J. J. Sakurai, Phys. Rev. D \textbf{21}, 2993 (1980).

[18] K.K. Nandi and Y.-Z. Zhang, Phys. Rev. D \textbf{66}, 063005 (2002).
\qquad

[19] P.M. Alsing, J.C. Evans and K.K. Nandi, Gen. Rel. Grav. \textbf{33},
1459 (2001).

[20] J. M. Cohen and B. Mashhoon, Phys. Lett. A \textbf{181}, 353 (1993).

[21] H.I.M. Lichtenegger and L. Iorio, Eur. Phys. J. Plus \textbf{126}, 129
(2011).

[22] C.W.F. Everitt \textit{et al.} Phys. Rev. Lett. \textbf{106}, 221101
(2011); Class. Quantum Grav. \textbf{32}, 224001 (2015).

[23] D.E. Smith and P.J. Dunn, Geophys. Res. Lett. 7, 437 (1980).

[24] M.L. Ruggiero,\ Phys. Rev. D \textbf{79}, 084001 (2009)

[25] E. Hackmann and C. L\"{a}mmerzahl, Phys. Rev. D \textbf{85}, 044049
(2012) and references therein.

[26] E. Hackmann and C. L\"{a}mmerzahl, Phys. Rev. Lett. \textbf{100},
171101 (2008).

[27] E. Hackmann, V. Kagramanova, J. Kunz and C. L\"{a}mmerzahl, Phys. Rev.
D \textbf{81}, 044020 (2010).

[28] A.D. Dolgov and M. Kawasaki, Phys. Lett. B \textbf{573}, 1 (2003).

[29] V. Kagramanova, J. Kunz and C. L\"{a}mmerzahl, Phys. Lett. B \textbf{634%
}, 465 (2006).

[30] M. Sereno and P. Jetzer, Phys. Rev. D \textbf{73}, 063004 (2006).

\begin{center}
\bigskip .\qquad -----------------------------------------------------
\end{center}

\end{document}